# Correlated electronic states of SrVO$_3$ revealed by angle-resolved photoemission spectroscopy

T. Yoshida[1], M. Kobayashi[2], K. Yoshimatsu[2,3], H. Kumigashira[2], A. Fujimori[3]

[1]Graduate School of Human and Environmental Studies, Kyoto University, Sakyo-ku, Kyoto 606-8501, Japan

[2]KEK, Photon Factory, Tsukuba, Ibaraki 305-0801, Japan

[3]Department of Physics, The University of Tokyo, Tokyo 113-0033, Japan

Abstract

In this article, we review recent progress in angle-resolved photoemission (ARPES) studies of the Mott-Hubbard-type correlated electron systems SrVO$_3$. It has the $d^1$ electron configuration and is an ideal model compound to study electron correlation effects in normal metal. ARPES studies of bulk single-crystal SrVO$_3$ and CaVO$_3$ have revealed the difference in the mass renormalization of electrons between them. *In-situ* ARPES studies of thin films fabricated by the pulsed laser deposition method have clarified not only quasi-particle dispersions, which show a kink like high-$T_c$ cuprates, but also finite dispersions in the incoherent part. Self-energy in a wide energy range has been deduced from the ARPES spectral weight using Kramers-Kronig transformation. The obtained self-energy has several structures which yield the incoherent feature and a pseudogap-like dip similar to the high-$T_c$ cuprates. Quantum-well states in ultrathin films of SrVO$_3$ have revealed sub-bands with correlated electrons. These findings of electron correlation effects outlined in the present article would provide a starting point not only for fundamental condensed-matter physics but also for the development of new devices with correlated electrons.

## 1. Introduction

Mott transitions, namely, metal-insulator transitions driven by electron correlation, have been the subject of extensively studies because they are outstanding phenomena in condensed matter [1]. According to dynamical mean-field theory (DMFT) calculations of Mott-Hubbard systems, incoherent spectral features develop, which is reminiscent of the lower and upper Hubbard bands, representing the localization of electrons [2,3]. At the same time, the bandwidth of quasiparticles crossing the Fermi level ($E_F$) becomes narrower with increasing $U/W$, i. e., the quasiparticle band is more strongly renormalized, where $U$ and $W$ are the on-site Coulomb repulsion and the one-electron bandwidth, respectively. The narrowing of the bandwidth has been theoretically predicted for a correlated Fermi liquid. However, the metal-insulator transition from the Fermi-liquid side has been far from well understood. As a prototype of the Fermi liquid, the perovskite-type oxide SrVO$_3$, which has the $d^1$ electron configuration, has been extensively investigated by photoemission spectroscopy [4,5,6] and also by theoretical calculations such as dynamical mean field theory (DMFT) [7]. In the early photoemission studies [4,5,6], as predicted by the DMFT calculation [3], so-called "incoherent part" has been observed at ~ -1.5 eV below the $E_F$ corresponding to the lower Hubbard band.

Narrowing of the quasiparticle band (coherent part) with increasing $U/W$ has also been observed to some extent consistent with the DMFT calculation [3]. However, whether there is difference in the mass renormalization between $SrVO_3$ and $CaVO_3$ or not has been controversial. Since $CaVO_3$ is orthorhombically distorted, it is expected to have a smaller $W$ and hence to show stronger mass renormalization than $SrVO_3$. Early photoemission results have shown that a dramatic suppression of the spectral weight of the coherent part has been observed in going from $SrVO_3$ to $CaVO_3$ but without appreciable band narrowing [6]. In contrast, according to a bulk-sensitive photoemission study using soft x-rays, neither appreciable spectral weight transfer nor band narrowing has been observed between $SrVO_3$ and $CaVO_3$ [8]. On the other hand, another bulk-sensitive photoemission study using a laser has revealed the suppression of spectral weight near $E_F$ in going from $SrVO_3$ to $CaVO_3$ [9].

The differences between the different studies may stem from the difference in the probing depth of photoemission spectroscopy due to the different photon energies. It has been pointed out by Sarma and co-workers [10,11] that the incoherent part for $SrVO_3$ and $CaVO_3$ has a lot of contributions from surface states through photoemission studies using different photon energies. In fact, a recent hard x-ray photoemission study revealed enhanced coherent spectral weight due to the bulk sensitivity [12]. Therefore, one should be careful when the spectral weight transfer and the effect of electron correlation are analyzed quantitatively. To reconcile the different results from the different experiments, angle-resolved photoemission spectroscopy (ARPES) turned out to be a powerful method. In a previous study, we succeeded in the ARPES observation of band dispersions and Fermi surfaces in $SrVO_3$ for the first time [13]. Since then, ARPES studies of the Mott Hubbard system have made a remarkable progress including *in situ* studies of thin film samples [14, 15, 16]. In this paper, we review recent ARPES studies of $SrVO_3$ and $CaVO_3$. The organization of this article is as follows: Section 2 describes the mass renormalization of bulk $SrVO_3$ and $CaVO_3$ [17], Sections 3 and 4 the quasi-particle and the self-energy of thin film $SrVO_3$ [14,16], and Section 5 the observation of quantum well states in $SrVO_3$ thin films [15, 18].

## 2. ARPES study of bulk single-crystal $SrVO_3$ and $CaVO_3$

First, let us overview the local-density-approximation (LDA) band-structure calculation of $SrVO_3$ [19,20]. While $SrVO_3$ has one electron in the V $3d$ orbitals, the three $t_{2g}$ bands are almost degenerate. As shown in Fig. 1(a), the three $t_{2g}$ bands consist of atomic orbitals $d_{xy}$, $d_{yz}$ and $d_{zx}$, respectively. Because hybridizations between them are weak and hence each band has nearly single orbital character, each of the $d_{xy}$, $d_{yz}$ and $d_{zx}$ orbitals forms a nearly two dimensional band with a cylindrical Fermi surface. In Fig. 1(b), three cylindrical Fermi surfaces penetrate each other and form three sheets. Such cylindrical Fermi surfaces have also been confirmed for $CaVO_3$ by de Haas van Alphen measurements [22].

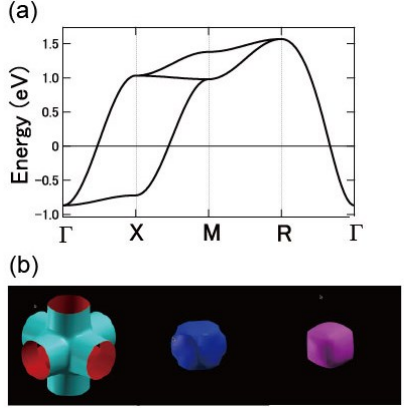

Figure 1: Calculated band structure of $SrVO_3$. (a) Three band dispersions for the $t_{2g}$ orbital ($xy$, $yz$, $zx$) [19]. (b) Fermi surfaces originating from the three two-dimensional Fermi surfaces consisting of the three $t_{2g}$ orbitals [21].

Yoshida *et al*. [17] have performed ARPES measurements of $SrVO_3$ and $CaVO_3$ on cleaved bulk single crystals. Fermi surface mapping in the $k_x$-$k_y$ plane is shown in Fig. 2(a). A two-dimensional circular $d_{xy}$ Fermi surface is observed. $d_{yz}$ and $d_{zx}$ Fermi surfaces, which are extended along the $k_x$ and $k_y$ directions, respectively, are also observed, consistent with the result of the band-structure calculation shown in the inset [20]. This indicates that they have nearly two-dimensional Fermi surfaces as expected from the band-structure calculation in spite of their three-dimensional crystal structures.

ARPES spectra of $SrVO_3$ along the Γ–X line ($k_x$ direction) of the Brillouin zone are shown in Figs. 2(b) and 2(c). The spectra within 0.7 eV of the $E_F$ constitute the coherent part and show clear dispersive features corresponding to the calculated band structure. The bottom of the coherent band is located at ~ 0.5 eV below $E_F$, which is nearly half of that predicted from the LDA calculation, corresponding to the factor of ~2 enhancement of the electronic specific heat coefficient γ [23]. This indicates that, while the shapes of the Fermi surfaces are almost the same as the prediction by the LDA calculation, there is a clear electron mass renormalization with a factor of ~2. Therefore, it is reasonable to understand the electronic structure of $SrVO_3$ as a Fermi liquid with an enhanced effective mass. On the other hand, the incoherent part, which is created by electron correlation, is centered at ~ -1.5 eV and shows appreciable momentum-dependent intensities while the dispersions are weak compared to the coherent part.

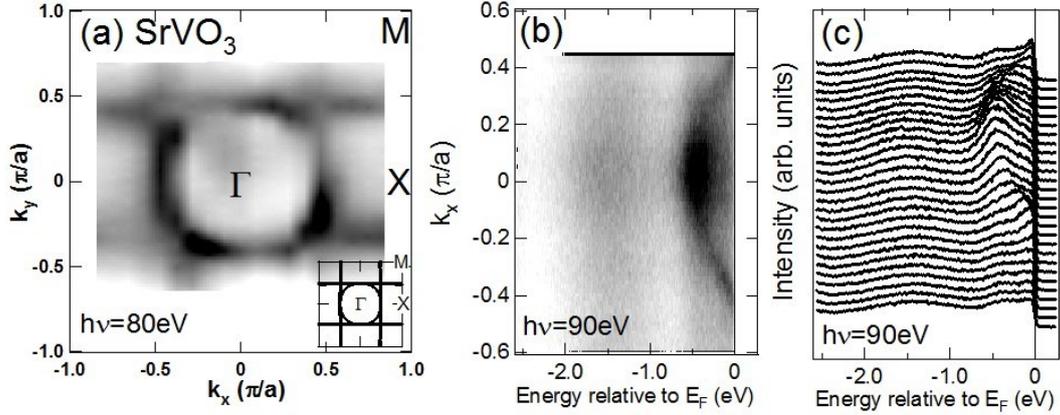

Figure 2: ARPES spectra of bulk $SrVO_3$ [17] (a) Fermi surface mapping. (inset) Fermi-surface cross sections on the $k_z=0$ plane predicted by band-structure calculation [20]. (b) Intensity plot in the $E$-$k_x$ plane with $k_y=0$. (c) Corresponding energy distribution curves (EDCs). The dispersive feature within 0.7 eV of $E_F$ is the coherent part while the broad feature centered around -1.5 eV below $E_F$ is the incoherent part.

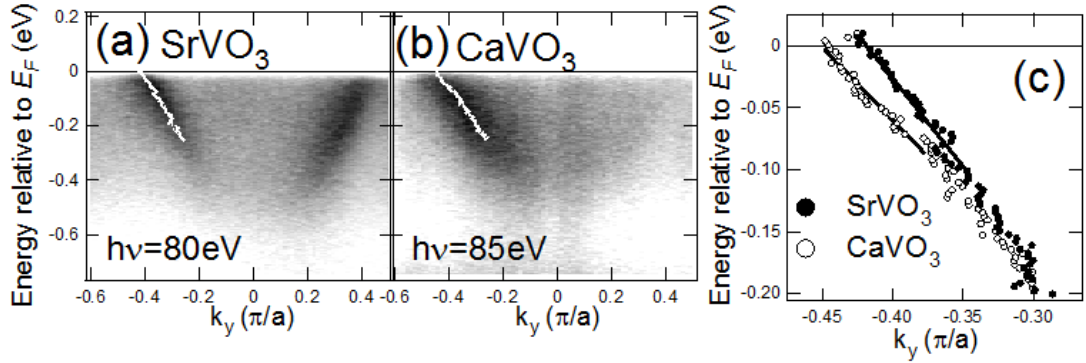

Figure 3: Comparison of the $d_{xy}$ band dispersion in $SrVO_3$ and $CaVO_3$ [17]. ARPES spectra of (a) $SrVO_3$ and (b) $CaVO_3$ along the Γ-X lines. (c) Comparison of the quasiparticle dispersions determined by the MDC peak positions. $CaVO_3$ has a smaller Fermi velocity than $SrVO_3$.

In $CaVO_3$, since the orthorhombic distortion causes band narrowing, one may expect that $U/W$ is larger and therefore that the electron correlation effects is stronger than that of $SrVO_3$. However, this prediction has not been unambiguously confirmed yet because of the surface sensitivity of photoemission spectroscopy, as mentioned in the introduction. In Fig. 3, we show direct comparison of the band dispersion of $SrVO_3$ and that of $CaVO_3$ [17]. As shown in Fig. 3(a) and 3(b), quasi-particle dispersions have been determined by MDC peak positions for both samples. Through detailed comparison between the quasiparticle (QP) bands in panels (a) and (b), it has been revealed that the Fermi velocity in $CaVO_3$ is smaller by ~20 % than that in $SrVO_3$. This band narrowing may be caused by the orthorhombic distortion and/or the increase of the electron correlation strength, i.e., the increase of $U/W$. While the soft x-ray photoemission study by Sekiyama

*et al*. [8] has shown that the widths of the coherent bands are nearly the same between both samples, ARPES with improved energy resolution was necessary to detect the finite difference in the Fermi velocity between CaVO$_3$ and SrVO$_3$.

## 3. *In-situ* ARPES study of thin film SrVO$_3$

Recently, more detailed ARPES measurements have become possible by using high-quality single-crystalline thin films having atomically flat surfaces grown by the pulsed laser deposition (PLD) technique [14,16]. Takizawa *et al*. [14] studied the electronic structure of SrVO$_3$ thin films by *in situ* ARPES measurements. Clear band dispersions were observed not only in the coherent QP part but also in the incoherent part, as shown in Fig. 4. A weak but finite (~0.1eV) dispersion of the incoherent part is seen in the momentum region enclosed by the Fermi surfaces. This is well reproduced by the DMFT calculation as shown in Fig. 4(b). This result indicates that part of the spectral weight of the coherent QP dispersion is transferred to the incoherent part and hence that the incoherent Hubbard band is influenced by the electron hopping between atoms.

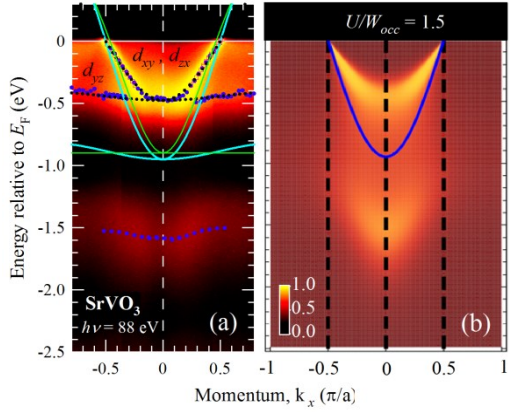

Figure 4: ARPES spectra of SrVO$_3$ thin film. (a) Experimental intensity plot. Peak positions of the EDCs are shown by blue filled circles. Blue curves are the V 3*d* bands from the LDA band structure. Broken curves are LDA bands renormalized by a factor of 2. The accuracy of the Fermi momentum is about 10% in this measurement. (b) Intensity plot of the spectral function from the DMFT calculation with $U/W_{occ}$=1.5, where $W_{occ}$=0.4$W$ is the width of the occupied part of the *d* band.

Aizaki *et al*. [16] have performed a self-energy analysis using the clear QP dispersion near $E_F$ of the thin film SrVO$_3$. In order to deduce the real part of the self-energy Re$\Sigma$, following the analysis frequently performed for the high-$T_c$ cuprates, under the assumption that the LDA result gives the non-interacting band dispersion, they have taken the difference between the band dispersions determined by the MDC peaks and the LDA band [20] as shown in Fig. 5(a). In the deduced Re$\Sigma$ shown in Fig. 5(b), a kink is seen around 60 meV below $E_F$ as indicated by an arrow

[24], which is very similar to those observed in the high-$T_c$ cuprates. In the studies of the high-$T_c$ cuprates, electron-phonon interaction [24], antiferromagnetic fluctuations, and/or the magnetic resonance mode [25] have been proposed as possible origins of the kink ~60 meV. In the case of SrVO$_3$, low-energy spin fluctuations are absent unlike the cuprates. In the perovskite-type compounds, Jahn-Teller phonons and oxygen breathing phonons are always observed in the energy range of 60–90 meV [26,27]. Therefore, the kink observed in the Re$\Sigma$ should be due to a coupling of electrons with these phonon modes characteristic of the perovskite oxides.

To determine the QP band dispersion near the band bottom, one should use the EDC peaks instead of MDC peaks as shown in Fig. 5(b). Here, it should be noted that even if we use the EDC peaks, the energy range of the deduced Re$\Sigma$ is limited to the bottom of the QP band (~ 0.5 eV below $E_F$), while the self-energy over a wider energy range is necessary to understand the electron correlation phenomena including the appearance of the remnant of the Hubbard bands. In the next section, we shall show a method to deduce the self-energy in a wider energy range including the region below the band bottom.

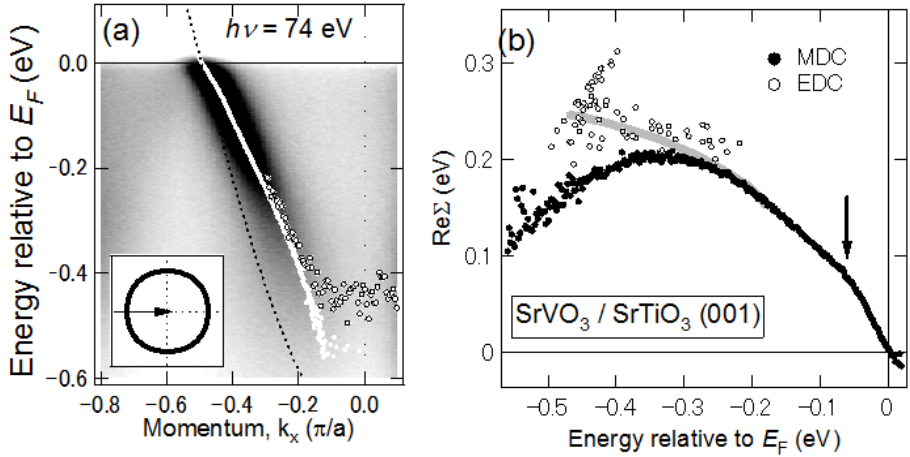

Figure 5: Band dispersions and self-energies for thin film SrVO$_3$ [16]. (a) $E$-$k$ intensity plot for a momentum cut shown in the inset. The QP band dispersions are determined by MDC peak positions (filled white circles), which are correct near $E_F$, and the second derivatives of EDCs (open circles), which are correct near the band bottom. The accuracy of the Fermi momentum is about 10% in this measurement. The non-interacting band given by the LDA band-structure calculation is shown by a broken curve. (b) Real part of the self-energy Re$\Sigma$. Here, we have obtained Re$\Sigma$ as the difference between the experimental band dispersion and the LDA energy band dispersion. The position of the kink at ~ 60 meV below $E_F$ is shown by an arrow.

## 4. Self-energy analysis

The energy range of the self-energy $\Sigma$ studied by the conventional method mentioned above is limited from $E_F$ to ~ -0.5 eV below it, that is, the energy range of the coherent part, while the behavior of the self-energy over a wider energy range is necessary to understand the role of electron correlation including the incoherent part. A phenomenological analytical function which fulfills Kramers-Kronig (KK) relation has often been introduced as a self-energy to reproduce both the shape of coherent and incoherent part [5,6]. However, such a form of the analytical function needs to be justified. In order to overcome this problem, Aizaki *et al.*, [16] have developed a new method to deduce the self-energy in a wider energy range starting from the spectral function itself and utilizing the Kramers-Kronig (KK) relation as described below.

Since the spectral function $A(k,\omega)$ is given by $(-1/\pi)\mathrm{Im}G(k,\omega)$, one can in principle obtain $\mathrm{Re}G(k,\omega)$ through the KK transformation of $\mathrm{Im}G(k,\omega)$, and consequently $G(k,\omega)=1/(\omega-\varepsilon_k-\Sigma(k,\omega))$. However, ARPES only gives $A(k,\omega)$ below $E_F$, i.e., $\omega<0$. Therefore, in this analysis, they assumed electron-hole symmetry for the self-energy. To begin with, one can construct the initial function for $A(k,\omega)$ as $A(k,\omega)= I(k,\omega)$ (for $k < k_F$) and $A(k,\omega)=I(k,\omega)+I(k,-\omega)$ (for $k=k_F$). By KK transformation, $G(k,\omega)$ and $\varepsilon_k+\Sigma(k,\omega)$ are obtained. In order to fulfill the electron-hole symmetry ($\mathrm{Re}\Sigma(k,\omega)= -\mathrm{Re}\Sigma(k,-\omega)$, $\mathrm{Im}\Sigma(k,\omega)= \mathrm{Im}\Sigma(k,-\omega)$), the self-energy $\Sigma(k,\omega)$ for $\omega>0$ can be renewed by using $\Sigma(k,\omega)$ for $\omega<0$. Thus obtained new $\Sigma(k,\omega)$ gives new $A(k,\omega)$. This cycle is iterated until these functions are converged.

According to this analysis, $A(k,\omega)$ including $\omega>0$ and momentum-averaged self-energy $\Sigma(\omega)$ have been obtained in a wide energy range as shown in Figs. 6(a) and 6(b). The experimental self-energy qualitatively reproduces characteristic features of the LDA+DMFT calculation [7]. Re$\Sigma$ shows a maximum at ~ -0.7 eV, where the EDC shows an intensity minimum, and this energy scale may also be related to the incoherent peak at ~ -1.5eV. Furthermore, the non-interacting band dispersion $\varepsilon_k$ has been deduced experimentally through $\mathrm{Re}[-1/G(k,0)]=\varepsilon_k$ without the knowledge of the LDA band structure [Fig. 6(c)]. Thus deduced $\varepsilon_k$ is in good agreement with the band-structure calculation [20]. The above recipe has successfully deduced the self-energy over the energy range including the incoherent part, and would be useful for future studies of electron correlation effects.

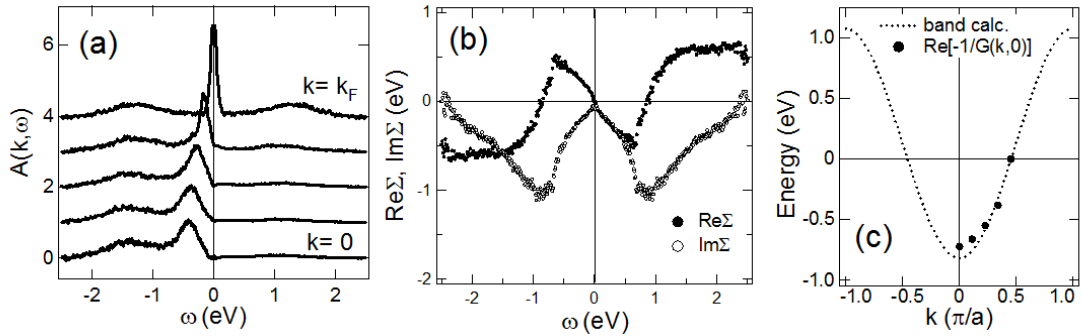

Figure 6: Self-energy iteratively deduced from the measured ARPES spectra utilizing the Kramers-Kronig transformation. (a): Self-consistently obtained $A(k,\omega)$ . (b) Self-consistently obtained Re$\Sigma$ and Im$\Sigma$ averaged over momentum. (c) Comparison of the self-consistently obtained Re[-1/$G(k,0)$] with the LDA band dispersion [20].

We have further analyzed characteristic features of the self-energy Σ [Fig. 6(b)] as follows. We have fitted the self-consistent self-energy to a phenomenological function $\Sigma(\omega) = \sum_n a_n \left( \frac{1}{\omega - \varepsilon_n + i\gamma_n} + \frac{1}{\omega + \varepsilon_n + i\gamma_n} \right)$, whose real part and imaginary part satisfy the KK relation. Figure 7(a) shows the fitted self-energy where the γ parameters have been reduced in order to emphasize spectral features. Figures 7(b) and 7(c) are $A(k,\omega)$=-Im$G(k,\omega)/\pi$ and Re$G(k,\omega)$ for $k$=0 calculated using the self-energy in Fig. 7(a). To identify poles and zeros of the Green function, we have drawn a line $\omega$-$\varepsilon_k$ in Fig. 7(a) and found three crossing points between ReΣ(ω) and the line. The crossing points at ~ -0.4 eV and ~ -1.5 eV yield poles of the Green function and corresponding to the coherent and incoherent peaks in $A(k,\omega)$, respectively. On the other hand, the cross point at ~ -0.8 eV is not a pole but gives a point where Re$G$=0, yielding a dip in $A(k,\omega)$ as shown in Fig. 7(b). In the analysis of the electronic structure of the high-$T_c$ superconductors, it has been discussed that a momentum region around Re$G$=0 gives a pseudogap in the spectral function [28]. In the case of SrVO$_3$, "the zero" gives a dip below (and above) $E_F$ which divides the spectral function into the coherent and incoherent part, in contrast to the high-$T_c$ cuprates which has "the zero" at $E_F$ yielding a pseudogap [28]. Although "the zero" in SrVO$_3$ has a different character from the high-$T_c$ cuprates, there is an analogy between them at least in mathematical structure of the self-energy. Therefore, further analysis of the self-energy of the Fermi liquid would be useful for the studies of correlated electron systems.

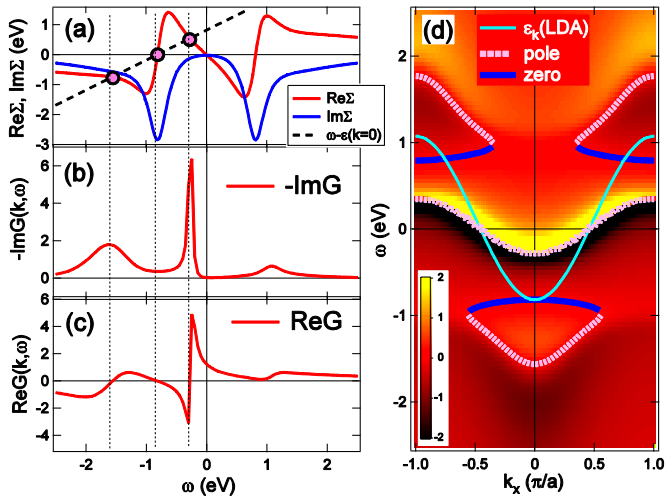

Figure 7: Simulation of the self-energy and the zero surface [Re$G(k,\omega)$=0]. (a) Self-energy Σ(ω) self-consistently deduced from the experimental data. (b),(c) Green function at $k$= 0, Im$G(0,\omega)$ =-π$A(0,\omega)$ and Re$G(0,\omega)$. (d) Image plot of Re$G$. The loci of its pole and the zero surface in energy-momentum space.

## 5. Quantum well

In contrast to the bulk electronic structure discussed above, novel electronic structures distinct from that of bulk often appear in ultrathin films. When the thickness of SrVO$_3$ film synthesized by PLD is less than ~10 monolayer, quantum well (QW) states are observed [15,29]. Because of the quantum confinement of V 3$d$ electrons in the two-dimensional space, the bands derived from the $d_{zx}$ and $d_{yz}$ orbitals, which are extended towards the z direction, form quantized electronic states as shown in Fig. 8, i.e., orbital-selective quantization. While the quantized bands are similar to those of the two dimensional electron gas (2DEG) observed at the surfaces and interfaces of oxide semiconductors, the characteristic electron density (~$10^{22}$cm$^{-3}$) and the size of potential well (~ few nm) are very different from those of the 2DEG systems (~$10^{18}$-$10^{20}$ cm$^{-3}$ and ~several nm) [30, 31].

Recently, it has been revealed that the effective electron mass in the quantized subbands depends on the energy of each band bottom (Figure 9) [15,32]. As the bottom of the subband approaches $E_F$, anomalous enhancement of the effective electron mass, which is a few times larger than that in the bulk band, has been observed [15]. Since such a strong mass enhancement has not been observed in typical QW states in metal thin films, this can be viewed as a characteristic phenomenon originating from strong electron correlation between $d$ electrons which are confined in quasi-one-dimensional space [18]. Fundamental understanding of the QW states with correlated electrons in SrVO$_3$ would provide a new way to material design in functional oxide devices by using the QW states.

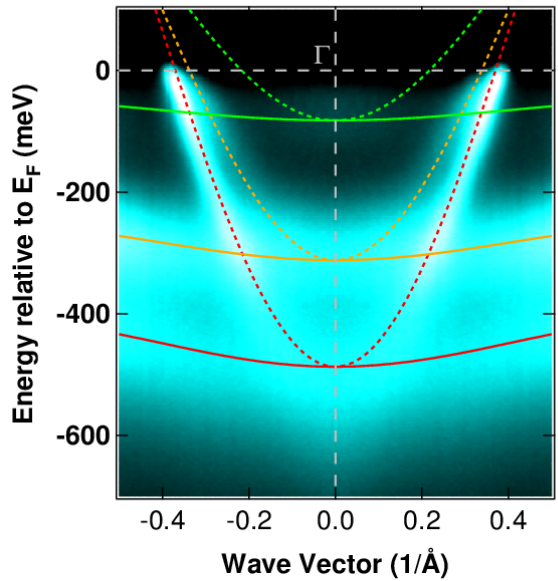

Figure 8. Intensity plot of ARPES spectra for a 7-ML SrVO$_3$ ultrathin film for cut along the $k_x$ direction across the Γ point. The intensity plot is symmetrized with respect to the center line and averaged. The dashed and solid curves

present the fitted curves to the dispersion using the tight-binding calculation for the $d_{zx}$ and $d_{yz}$ quantized states, respectively. Red, orange, and green curves correspond to state in which $n$ = 1, 2, and 3, respectively [15].

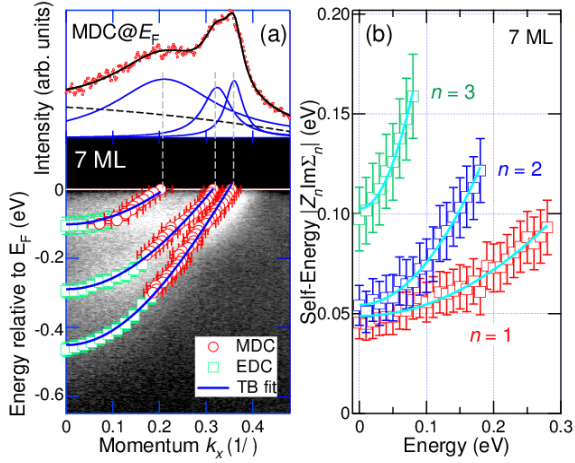

Figure 9. Line-shape analysis of ARPES data for a 7-ML SrVO$_3$ ultrathin film [18]. (a) (bottom) ARPES image. The open circles and squares are peak positions estimated from the MDCs and EDCs, respectively. The solid curves represent the fit to tight-binding calculations. (top) MDC at $E_F$. The curve is fitted to superposed Lorentzians. The dashed curve is a smooth background. (b) $\omega$-dependence of the imaginary part of the self-energy $|Z_n\mathrm{Im}\Sigma_n|$. The solid curves represent the fit to a parabolic function $\omega^2$.

## 6. Summary

From the photoemission studies overviewed in this article, we have demonstrated that SrVO$_3$, which has the $d^1$ electron configuration, is an ideal model compound for the studies of electron correlation effects in correlated Fermi liquids. In the ARPES studies of bulk single crystals of SrVO$_3$ and CaVO$_3$, the difference in the mass renormalization between them has been identified. Thin films fabricated by the PLD method have high-quality surfaces and make it possible to investigate the detailed QP properties including the incoherent part. Particularly, the self-energy of an electron has been deduced in a wide energy range from the ARPES spectra by utilizing the Kramers-Kronig transformation. The self-consistently obtained self-energy exhibits a pseudogap-like feature similar to the high-$T_c$ cuprates between the coherent and incoherent spectral features. This analogy to the pseudogap of the cuprates may be applied to other correlated system in the future. It has been clarified that ultrathin films of SrVO$_3$ show QW states and their sub-bands also exhibit electron correlation effects. Since it has been pointed out that the spectral weight of the incoherent part has contributions not only from the V 3$d$ orbital but also from the O 2$p$ orbital, the role of the O 2$p$ orbital may have to be taken into account to fully understand the electron correlation effect in SrVO$_3$ [33,34]. By constituting a fundamental understanding of electron correlation in bulk and thin film SrVO$_3$, the development of new devices with correlated electrons would be expected in artificially synthesized oxide superlattices and hetero-structures.

**Acknowledgement**

We would like to thank fruitful and enlightening discussion with D. D. Sarma, K. Maiti, S. Shin, H. Eisaki, J. C. Fuggle, O. K. Andersen, S. Biermann, A. Georges, M. Imada, M. J. Rozenberg, and S. Okamoto. This work is supported by Grants-in-Aid for Scientific Research (15H02109, B25287095 and S22224005) and for Young Scientists (268708423) from JSPS, Elements Strategy Initiative to Form Core Research Center from the MEXT, and a grant from The Murata Science Foundation. This work was done under the approval of the Photon Factory Program Advisory Committee (Proposals No. 2007G597, 2008S2-003, 2013G218, and 2015G144).